\documentstyle[12pt]{article}

\newcommand{\beq}{\begin{equation}}
\newcommand{\eeq}{\end{equation}}
\newcommand{\beqa}{\begin{eqnarray}}
\newcommand{\eeqa}{\end{eqnarray}}
\newcommand{\beqar}{\begin{eqnarray*}}
\newcommand{\eeqar}{\end{eqnarray*}}

\newcommand{\Ga}{\Gamma}

\newcommand{\inn}{\!\cdot\!}

\newcommand{\z}{\zeta}

\newcommand{\eg}{{\it e.g.,}\ }
\newcommand{\ie}{{\it i.e.,}\ }
\newcommand{\labell}[1]{\label{#1}} %{\label{#1}} %
\newcommand{\reef}[1]{(\ref{#1})}
\newcommand\prt{\partial}

\newcommand\tG{{\widetilde G}}

\newcommand\tV{{\widetilde V}}

\begin{document}

\thispagestyle{empty} \rightline{\small hep-th/0303239 \hfill
IPM/P-2003/014} \vspace*{1cm}

\begin{center}
{\bf \Large
Off-shell extension of  S-matrix elements\\
and tachyonic effective actions  \\

 }
\vspace*{1cm}

{Mohammad R. Garousi}\\
\vspace*{0.2cm}
{\it Department of Physics, Ferdowsi university, Mashhad, Iran}\\
\vspace*{0.1cm}
and\\
{\it Institute for Studies in Theoretical Physics and Mathematics
IPM} \\
{P.O. Box 19395-5531, Tehran, Iran}\\
\vspace*{0.4cm}

\vspace{2cm} ABSTRACT
\end{center}
We show that the on-shell S-matrix elements of four open string
massless scalars, two scalars and two tachyons, and four open
string  tachyons in the super string theory can be written in a
unique form. We then propose an off-shell extension for the
S-matrix element of four scalars which is consistent, in the low
energy limit,  with the Dirac-Born-Infeld effective action. Using
a similar off-shell extension for the S-matrix element of two
scalars and two tachyons and for the S-matrix element of four
tachyons, we show that they are fully consistent with the
tachyonic DBI action. \vfill \setcounter{page}{0}
\setcounter{footnote}{0}
\newpage

\section{The idea} \label{intro}
Recently different tachyonic effective actions has been used to
describe the time evolution of unstable D-branes in string
theory\cite{sen1,gg,ssst,jm,nlis}\footnote{For early studies of
open string tachyon dynamics, see \cite{kb}.}. In particular, Sen
has shown that the string theory produces a pressure-less gas with
non-zero energy density at the late time of the tachyon
condensation\cite{sen1}. In this paper, he  showed that these
results can be derived also from the tachyonic
Dirac-Born-Infeld(DBI) effective action\cite{mg,ebmr} around its
true vacuum. Other possible applications of this action to
cosmology have been discussed in \cite{cosmotachyon}. The
tachyonic DBI action was first found  by analyzing  the low energy
behaviour of the amplitude describing scattering of two tachyons
to one graviton on the noncommutative world-volume of a non-BPS
D$_p$-branes in superstring theory \cite{mg}. Then it was shown
that the action is consistent with T-duality rules, and has
expected solitonic solutions\cite{ebmr}.

In inferring effective action from on-shell  string theory
S-matrix elements, one usually evaluates the  S-matrix elements
and  expands them in the limit $\alpha'\rightarrow 0$, \ie low
energy limit. Then one writes a low energy effective action in
the field theory that its corresponding on-shell S-matrix elements
reproduce the leading order terms of the expansion \cite{js}. The
low energy expansion of the on-shell S-matrix elements are
unique, however, many apparently different but physically
identical low energy actions may produce them. These actions may
in turn be
 related to each other by some field
redefinition \cite{rmat}. Alternatively, one may try to extend
appropriately   the on-shell S-matrix elements to different
off-shell amplitudes and then expand them in the low energy limit.
Each amplitude  can be reproduced in field theory by a unique
action. These actions should then be related to each other by some
field redefinition.   For example, consider the S-matrix element
of one closed string tachyon and two open string tachyons in Type
0 theory \cite{ahik,mg}:
 \beqa
A(\tau,T,T)&\sim&\left(\frac{\Ga(-2s)}{\Ga(-s)\Ga(-s)}\right)\,\,,
\label{att}\eeqa where $\tau$ and $T$ stand for the closed string
and open string tachyon, respectively. Also  $s=-\alpha'(k_1+
k_2)^2/2$, and $k_1^a,k_2^a$ are momenta of the open string
tachyons which satisfy the on-shell condition
$k_1^2=k^2=1/(2\alpha')$. One may analytically off-shell extends
this amplitude to \beqa A^{\rm off}
(\tau,T,T)&\sim&\left(\frac{\Ga(-2s)}{\Ga(-s)\Ga(-s)}
+1/4-\alpha'(k_1^2+k_2^2)/4+f(k_1, k_2)\right)\,\,,
\labell{att1}\eeqa where the tachyon momenta do not satisfy the
on-shell condition anymore, and the function $f(k_1,k_2)$, having
only some contact terms of order $O(\alpha'^2)$, must be
symmetric under $1\leftrightarrow 2$ and vanishes on-shell.
 Obviously  the off-shell amplitude \reef{att1} reduces to the
on-shell amplitude \reef{att} upon imposing  the on-shell
condition $k_1^2=k_2^2=1/(2\alpha')$. The low energy expansion,
\ie $\alpha'\rightarrow 0$, of \reef{att1} is \beqa A^{\rm
off}(\tau,T,T)&\sim&\left(1/4+\alpha'k_1\inn
k_2/2+O(\alpha'^2)\right)\,\,. \label{att2}\eeqa The first two
terms  are reproduced exactly by the tachyonic DBI
action\cite{mg,ebmr}\footnote{We have used the fact that the
closed string tachyon in Type 0 theory normalizes the D-brane
tension as $T_p\rightarrow T_p(1+\tau/4+\cdots)$\cite{mg3}, and
we have kept only linear term for the closed string tachyon.}
\beqa S&=&-\frac{T_p}{4}\int dx\,\tau
V(T)\sqrt{-\det(\eta_{ab}+2\pi\alpha'\prt_a T\prt_b T)}\,\,,
\label{dbiac}\\
&=&-\frac{T_p}{4}\int dx\,\tau\left(1-\pi T^2/2+\pi \alpha'(\prt
 T)^2+\cdots\right)\,\,,\nonumber\eeqa
 where in the second line we have used the
expansion $V(T)=1-\frac{\pi}{2}T^2+O(T^4)$ at the top of  the
tachyon potential in this action. Note that to evaluate the
off-shell S-matrix elements in field theory we do not use the
on-shell conditions for external states in Feynman diagrams,
however, we do use the conservation of momentum. The terms of
order $O(\alpha'^2)$ in \reef{att2} are related to higher
derivative terms which are not included in tachyonic action
above\footnote{The higher derivative terms may have significant
effect at the top of potential, however, according to the results
in \cite{sen1} they are not important at the minimum of tachyon
potential.}. Now consider the following  off-shell extension of
the  amplitude \reef{att}: \beqa A^{\rm
off}(\tau,T,T)&\sim&\frac{\Ga(1+2\alpha' k_1\inn
k_2)}{\Ga(1/2+\alpha' k_1\inn k_2)\Ga(1/2+\alpha' k_1\inn
k_2)}+g(k_1,k_2)\,\,,\nonumber\eeqa where function $g(k_1,k_2)$,
having only some contact terms of order $O(\alpha'^2)$, is zero
on-shell. Again it is obvious that this off-shell amplitude
reduces to the on-shell amplitude \reef{att} upon imposing the
on-shell condition. At low energy this amplitude has the following
expansion: \beqa A^{\rm off}&\sim& 1/4+\alpha'\ln(2)k_1\inn
k_2+O(\alpha'^2)\,\,.\labell{att3}\eeqa In this case, the first
two terms of this expansion are reproduced exactly by BSFT
action\cite{dkmm} \beqa S&=&-\frac{T_p}{4}\int dx \,\tau
e^{-T^2/4}\frac{\sqrt{\pi}\Ga(\alpha'(\prt
T)^2/2+1)}{\Ga(\alpha'(\prt T)^2/2+1/2)}\,\,,\labell{bsft}\\
&=&-\frac{T_p}{4}\int dx \,\tau\left(1-T^2/4+\alpha'\ln(2)(\prt
T)^2+\cdots\right)\,\,.\nonumber\eeqa Again the terms of order
$O(\alpha'^2)$ in \reef{att3} are related to the higher derivative
terms which are not included in the BSFT action. One expects that
the two actions \reef{dbiac} and \reef{bsft} to be related to
each other by some field redefinition\cite{pmas}.

Now using the idea that one can off-shell extend the S-matrix
elements, the question is how to off-shell extend a S-matrix
element? The particular case that we are interested in is how to
off-shell extend S-matrix elements involving open string
tachyons? The guiding principle  that we follow is the similarity
between on-shell S-matrix elements involving  massless scalars and
the on-shell S-matrix elements involving tachyons. We extend this
similarity to the off-shell amplitudes as well.  We off-shell
extend the amplitudes involving only scalars  in  such a way that
they are consistent with the DBI action in the low energy limit.
Then we use a similar off-shell amplitude for the cases that
involve tachyons. As a simple example consider the S-matrix
element of one closed string tachyon and two open string massless
scalar states. The amplitude is given by
\cite{mg2}\footnote{Using the fact that the closed strings are
not functional of the open string tachyon but are functional of
the scalar fields \cite{mg5}, we assumed that the closed string
field does not depend on the scalar. This makes similar the
S-matrix elements involving the scalars and the tachyon as much
as possible.}, \beqa A(\tau,X,X)&\sim&\z_1\inn
\z_2\left(\frac{\Ga(-2s)}{\Ga(-s)\Ga(-s)}\right)
\,\,,\labell{axxon}\eeqa where $s=-\alpha'(k_1+ k_2)^2/2$, and
$\z_1^i,\z_2^i$ and $k_1^a,k_2^a$ are the scalar polarizations
and momenta, respectively\footnote{Our index conventions are that
early Latin indices take values in the world-volume, \eg
$a,b=0,1,...,p$, and middle Latin indices take values in the
transverse space, \eg $i,j=p+1,...,8,9$.}. The momenta in this
amplitude satisfy the on-shell condition $k_1^2=k_2^2=0$. Note
that apart from the polarization of the scalar fields, the
on-shell amplitude \reef{att} and \reef{axxon} are similar.
Off-shell extension for the above amplitude which is correspond
to the DBI action is \beqa A^{\rm off}(\tau,X,X)&\sim&\z_1\inn
\z_2\left(\frac{\Ga(-2s)}{\Ga(-s)\Ga(-s)}+G-G^{{\rm on}} +
h(k_1,k_2)\right) \,\,,\labell{axxoff}\eeqa where the function
$G(k_1,k_2)$ is \beqa G&=&
-\alpha'(k_1^2+k_2^2)/4\,\,,\nonumber\eeqa and $G^{{\rm on}}$
means the function $G$ in which the momenta are on-shell.
Obviously it is zero in this case. The momenta in \reef{axxoff}
do not satisfy the on-shell conditions and the function
$h(k_1,k_2)$, having only some contact terms of order
$O(\alpha'^2)$, must be symmetric under $1\leftrightarrow 2$ and
must be zero on-shell. This function is related to the higher
derivative terms in the DBI action. The low energy expansion, \ie
$\alpha'\rightarrow 0$, of this amplitude is \beqa A^{\rm
off}&\sim&\z_1\inn \z_2\left(\alpha'k_1\inn k_2/2+O(\alpha'^2)
\right)\,\,.\nonumber\eeqa The first term above is reproduced in
field theory by the  DBI action as required, and $O(\alpha'^2)$
terms are related to the higher derivative terms which are not
included in the DBI action. Now the off-shell amplitude
\reef{att1} is similar to the off-shell amplitude \reef{axxoff},
\ie  \beqa A^{\rm off}
(\tau,T,T)&\sim&\left(\frac{\Ga(-2s)}{\Ga(-s)\Ga(-s)} +G-G^{{\rm
on}} +f(k_1, k_2)\right)\,\,. \labell{att11}\eeqa In this case
$G^{{\rm on}}=-1/4$. This off-shell amplitude is consistent with
the tachyonic DBI action as we saw above.  In the present paper we
would like to extend these idea to the case of S-matrix elements
of four tachyons and/or scalars in the superstring theory in the
presence of background B-flux.

In section 2, we recall the string theory S-matrix element of four
scalar vertex operators. We propose an off-shell extension for
this amplitude and then expand it  at low energy. We show that
its leading order terms are reproduced exactly by the
corresponding off-shell amplitude in field theory based on the DBI
action. In section 3, we show that   the on-shell S-matrix element
of two scalar and two tachyon vertex operators is similar to the
on-shell S-matrix element of four scalars. Hence we write an
off-shell S-matrix elements that is similar to the one in section
2. In  section 4 we do the same thing for the S-matrix element of
four tachyons. The proposed off-shell amplitudes in sections 3
and 4 are  correspond to the tachyonic DBI action.

\section{Four  scalars amplitude} Scattering amplitude of four
vector vertex operators in superstring theory is evaluated in
\cite{mgjs}, and its low energy effective action is also studied,
for example, in \cite{at}. To find the amplitude corresponding to
four scalar vertex operators, one may use the result in
\cite{mgjs} in which the vector polarizations $\z^a$ are replaced
by the scalar polarizations $\z^i$. Since we are interested in
the scattering amplitude in the presence of B-flux, one should
use $G=(1/(\eta+2\pi\alpha'B))_S$ as the world volume metric, and
also should add an appropriate  phase factor to the amplitudes in
one cycling of the vertex operators with the non-commutative
parameter tensor $\theta=(2\pi\alpha'/(\eta+2\pi\alpha'B))_A$
\cite{nsew}. Adding all non-cyclic permutation of the vertex
operators, one ends up with the following amplitude: \beqa
A(\z_1,\z_2,\z_3,\z_4)&=&\,A_s(\z_1,\z_2,\z_3,\z_4)+\,A_u
(\z_1,\z_2,\z_3,\z_4)+ \,A_t(\z_1,\z_2,\z_3,\z_4)\,\,,
\nonumber\eeqa where $A_s,A_u$, and $A_t$ are the part of the
amplitude that  has  massless pole in $s$-, $u$- and $t$-channels,
respectively. They are
 \beqa A_s&= &-\frac{icT_p}{2\pi^2\alpha'^2}\z_1\inn\z_2\,\z_3\inn\z_4
\left(\frac{}{}\frac{\Ga(-2s)\Ga(1-2t)}{\Ga(-2s-2t)}(e^{i\pi
l_{12}+i\pi l_{34}}
+e^{i\pi l_{14}-i\pi l_{23}})\right.\nonumber\\
&&+\frac{\Ga(-2s)\Ga(1-2u)}{\Ga(-2s-2u)}(e^{i\pi l_{13}-i\pi
l_{24}}+e^{i\pi
l_{12}-i\pi l_{34}})\nonumber\\
&&\left.-\frac{\Ga(1-2t)\Ga(1-2u)}{\Ga(1-2t-2u)}(e^{i\pi
l_{14}+i\pi l_{23}}+e^{i\pi l_{13}+i\pi l_{24}}) \frac{}{}
\right)\,\,,
\nonumber\\
A_u&=&-\frac{icT_p}{2\pi^2\alpha'^2}\z_1\inn\z_3\,
\z_2\inn\z_4\left(-\frac{}{}
\frac{\Ga(1-2s)\Ga(1-2t)}{\Ga(1-2s-2t)}(e^{i\pi l_{12}+i\pi
l_{34}}
+e^{i\pi l_{14}-i\pi l_{23}})\right.\nonumber\\
&&+\frac{\Ga(-2u)\Ga(1-2s)}{\Ga(-2u-2s)}(e^{i\pi l_{13}-i\pi
l_{24}}+e^{i\pi
l_{12}-i\pi l_{34}})\nonumber\\
&&\left.+\frac{\Ga(-2u)\Ga(1-2t)}{\Ga(-2u-2t)}(e^{i\pi l_{14}+i\pi
l_{23}}+e^{i\pi l_{13}+i\pi l_{24}}) \frac{}{}
\right)\,\,,\nonumber\\
A_t&=&-\frac{icT_p}{2\pi^2\alpha'^2}\z_1\inn\z_4\,
\z_2\inn\z_3\left(\frac{}{}
\frac{\Ga(-2t)\Ga(1-2s)}{\Ga(-2t-2s)}(e^{i\pi l_{12}+i\pi l_{34}}
+e^{i\pi l_{14}-i\pi l_{23}})\right.\nonumber\\
&&-\frac{\Ga(1-2s)\Ga(1-2u)}{\Ga(1-2s-2u)}(e^{i\pi l_{13}-i\pi
l_{24}}+e^{i\pi
l_{12}-i\pi l_{34}})\nonumber\\
&&\left.+\frac{\Ga(-2t)\Ga(1-2u)}{\Ga(-2t-2u)}(e^{i\pi l_{14}+i\pi
l_{23}}+e^{i\pi l_{13}+i\pi l_{24}}) \frac{}{}
\right)\,\,,\labell{aton}\eeqa where
$l_{\alpha\beta}=k_{\alpha}\inn \theta\inn k_{\beta}/(2\pi)$ for
$\alpha,\beta=1,2,3,4$, and variables $s,t,u$ are the
following:\beqa
s&=&-\alpha'(k_1+k_2)^2/2\,\,,\nonumber\\
t&=&-\alpha'(k_2+k_3)^2/2\,\,,\labell{mandel}
\\
u&=&-\alpha'(k_1+k_3)^2/2\,\,.\nonumber \eeqa Using on-shell
condition for the momenta, one finds that these Mandelstam
variables satisfy the on-shell relation $s+t+u=0$. We have also
normalized the amplitude by the factor $icT_p/(2\pi^2 \alpha'^2)$
where $c=\sqrt{-\det(\eta+B)}$.  The amplitudes $A_s$, $A_u$ and
$A_t$ are very similar, so we only off-shell extend the $A_t$
amplitude. The off-shell extension of $A_s$ and $A_u$ should then
be straightforward.

To find the off-shell extension of $A_t$ we use the following
criteria: \\
1)-The amplitude should have poles at $2t=0,1,2,\cdots$,
corresponding to propagation of  the infinite tower of open
string states in the amplitude.\\
2)-Imposing on-shell conditions, the amplitude must reduce to the
on-shell amplitude $A_t$.\\
3)-In the low energy expansion of the amplitude, its massless
pole and its contact terms up to $O(\alpha'^3)$ should produce
the corresponding pole and contact terms of the DBI field theory
amplitude.
\\
Unlike the amplitude \reef{axxon} in which the Mandelstam
variable $s$ is arbitrary, the Mandelstam variables in
\reef{aton} are constraint by the on-shell condition $s+t+u=0$.
Now it raises the question that how we write the gamma functions
in \reef{aton} in the off-shell amplitude. To do so, we first,
using the on-shell condition $s+t+u=0$,  write the on-shell
amplitude in a unique form that all other amplitudes involving the
tachyon  can also be converted to. It turns out that the gamma
function in $A_t$ should be written in the following form: \beqa
\frac{\Ga(-2t)\Ga(1-2s)}{\Ga(-2t-2s)}&=&
\frac{\Ga(-2t)\Ga(1+t+u-s)}{\Ga(u-s-t)}\,\,,\nonumber\\
\frac{\Ga(1-2s)\Ga(1-2u)}{\Ga(1-2s-2u)}&=&
\frac{\Ga(1+t+u-s)\Ga(1+s+t-u)}{\Ga(1+2t)}\,\,,\nonumber\\
\frac{\Ga(-2t)\Ga(1-2u)}{\Ga(-2t-2u)}&=&
\frac{\Ga(-2t)\Ga(1+t+s-u)}{\Ga(s-u-t)}\,\,.\labell{gamma1}\eeqa
Writing the gamma functions in \reef{aton} in the above form, we
propose the following off-shell amplitude for $A_t$:  \beqa
A_t^{\rm off}
&=&-\frac{icT_p}{2\pi^2\alpha'^2}\z_1\inn\z_4\,\z_2\inn\z_3\nonumber\\
&&\times\left(\left(\frac{}{}
\frac{\Ga(-2t)\Ga(1+t+u-s)}{\Ga(u-s-t)}+F-F^{{\rm on}}
+f\right)(e^{i\pi l_{12}+i\pi l_{34}}
+e^{i\pi l_{14}-i\pi l_{23}})\right.\nonumber\\
&&-\left(\frac{\Ga(1+t+u-s)\Ga(1+s+t-u)}{\Ga(1+2t)}-F+F^{{\rm on}}
-g\right) (e^{i\pi l_{13}-i\pi l_{24}}+e^{i\pi
l_{12}-i\pi l_{34}})\nonumber\\
&&\left.+\left(\frac{\Ga(-2t)\Ga(1+t+s-u)}{\Ga(s-u-t)}+F-F^{{\rm
on}} +h\right) (e^{i\pi l_{14}+i\pi l_{23}}+e^{i\pi l_{13}+i\pi
l_{24}}) \frac{}{} \right)\,\,,\labell{atoff}\eeqa where the
function $F(k_1,k_2,k_3,k_4)$ has been added to the amplitude to
cancel the non-desire contact terms resulting from expansion of
the gamma functions in the low energy limit. It has the following
contact terms of order $\alpha'^2$:\beqa
F&=&-\frac{\alpha'\pi^2}{6} \left(\alpha'(\sum_{i=1}^4 k_i^2)^2/4
+(k_1^2+k_4^2)t+(k_1^2+k_3^2)u+(k_1^2+k_2^2)s\right.\nonumber\\
&&\left.+\alpha'k_2\inn k_3(k_1^2+k_4^2)-\alpha'k_1\inn
k_3(k_2^2+k_4^2)-\alpha'k_1\inn
k_2(k_3^2+k_4^2)\right)\,\,.\labell{F}\eeqa $F^{{\rm on}}$ means
function $F$ in which the momenta are on-shell. It is easy to see
that this is zero in this case.   Functions
$f(k_1,k_2,k_3,k_4),g(k_1,k_2,k_3,k_4),h(k_1,k_2,k_3,k_4)$,
having only some contact terms of order $O(\alpha'^3)$, must be
zero on-shell. These functions are related to the higher
derivative corrections to the DBI action in which we are not
interested in the present paper. Using conservation of momentum,
the Mandelstam variables in the off-shell case satisfy the
relation: \beqa
s+t+u&=&-\alpha'(\sum_{i=1}^{4}k_i^2)/2\,\,.\labell{mandel1}\eeqa

Its is obvious that this off-shell amplitude satisfies the first
two criteria above. To check the last criterion, we should expand
the amplitude in the low energy. The gamma functions in this
amplitude at $\alpha'\rightarrow 0$ have the following expansion:
\beqa
\frac{\Ga(-2t)\Ga(1+t+u-s)}{\Ga(u-s-t)}&=&\frac{1}{2}+\frac{s-u}{2t}-
\frac{\pi^2}{6}\left((s-u)^2-t^2\right)+O(\alpha'^3)\,\,,\nonumber\\
\frac{\Ga(-2t)\Ga(1+t+s-u)}{\Ga(s-u-t)}&=&\frac{1}{2}+\frac{u-s}{2t}-
\frac{\pi^2}{6}\left((s-u)^2-t^2\right)+O(\alpha'^3)\,\,,\nonumber\\
\frac{\Ga(1+t+u-s)\Ga(1+s+t-u)}{\Ga(1+2t)}
&=&1+\frac{\pi^2}{6}\left((s-u)^2-t^2\right)+
O(\alpha'^3)\,\,.\labell{gamma}\eeqa  Replacing these expansion
for the gamma functions into \reef{atoff}, one finds massless
pole and some contact terms at each order of $\alpha'$.

Now in  field theory, using the non-commutative DBI action, one
finds the following massless
 $t$-channel amplitude (see \eg \cite{mg2} for details): \beqa
A_t'^{\rm off}
&=&(\tV_{\phi_2\phi_3A})^a(\tG_A)_{ab}(\tV_{A\phi_1\phi_4})
^b\,\,,\nonumber\\
&=&\left(\frac{icT_p}{(\alpha'\pi)^2}\right)\z_1\inn\z_4\,\z_2\inn\z_3
\,\frac{\sin(\pi l_{23})\sin(\pi
l_{14})\alpha'(k_2-k_3)\inn(k_1-k_4)}{2t}
\,\,,\nonumber\\
&=&\frac{-icT_p}{2\pi^2\alpha'^2}\z_1\inn\z_4\,\z_2\inn\z_3\,\left((e^
{i\pi l_{12}+i\pi l_{34}}+e^{i\pi l_{14}-i\pi
l_{23}})(\frac{s-u}{2t}
)\right.\nonumber\\
&&\left.\qquad\qquad\qquad\qquad+(e^{i\pi l_{14}+i\pi
l_{23}}+e^{i\pi l_{13}+i\pi l_{24}})(\frac{u-s}{2t}
)\right)\,\,.\labell{atfield}\eeqa In reaching to this result, we
have used only conservation of momentum. It easy to see that the
massless pole of the string theory off-shell amplitude
\reef{atoff} produces exactly the corresponding amplitude in the
field theory. Now subtracting the field theory massless pole from
the string theory  amplitude, one ends up with the following
contact terms in the $t$-channel: \beqa A_t^{\rm off}-A_t'^{\rm
off} &=&-\frac{icT_p}{2\pi^2\alpha'^2}\z_1\inn\z_4\,
\z_2\inn\z_3\labell{att'}\\
&&\times \left(\frac{}{}(e^{i\pi l_{12}+i\pi l_{34}} +e^{i\pi
l_{14}-i\pi
l_{23}})(\frac{1}{2}-\frac{\pi^2}{6}\left((s-u)^2-t^2\right)+F+O
(\alpha'^3)
)\right.\nonumber\\
&&+(e^{i\pi l_{13}-i\pi l_{24}}+e^{i\pi l_{12}-i\pi
l_{34}})(-1-\frac{\pi^2}{6}\left((s-u)^2-t^2\right)+F
+O(\alpha'^3))\nonumber\\
&&\left.+(e^{i\pi l_{14}+i\pi l_{23}}+e^{i\pi l_{13}+i\pi l_{24}})
(\frac{1}{2}-\frac{\pi^2}{6}\left((s-u)^2-t^2\right)+F+O(\alpha'^3)
)\frac{}{} \right)\,\,.\nonumber\eeqa Doing the same analysis for
the $A_s$ and $A_u$, one finds the following total off-shell
contact terms:\beqa A_c^{{\rm off}}
(\z_1,\z_2,\z_3,\z_4)&=&A_c^0(\z_1,\z_2,\z_3,\z_4)+
A_c^4(\z_1,\z_2,\z_3,\z_4)+
\sum_{n>4}A_c^n(\z_1,\z_2,\z_3,\z_4)\,\,,\nonumber\eeqa where
$A_c^0$ contains, apart from the phase factor, contact terms with
no momentum and is zero when the background B-flux vanishes, \beqa
A_c^0&\!\!=\!\!&\frac{-icT_p}{2\pi^2\alpha'^2}\left(e^{i\pi
l_{12}+i\pi l_{34}} +e^{i\pi l_{14}-i\pi
l_{23}}\right)\left(\frac{1}{2}\z_1\inn\z_4\,\z_2\inn\z_3
+\frac{1}{2}\z_1\inn\z_2\,\z_3\inn\z_4-\z_1\inn\z_3\,
\z_2\inn\z_4\right)\nonumber\\
&&-\frac{icT_p}{2\pi^2\alpha'^2}\left(e^{i\pi l_{13}-i\pi l_{24}}
+e^{i\pi l_{12}-i\pi
l_{34}}\right)\left(-\z_1\inn\z_4\,\z_2\inn\z_3
+\frac{1}{2}\z_1\inn\z_2\,\z_3\inn\z_4+
\frac{1}{2}\z_1\inn\z_3\,\z_2\inn\z_4\right)\nonumber\\
&&-\frac{icT_p}{2\pi^2\alpha'^2}\left(e^{i\pi l_{14}+i\pi l_{23}}
+e^{i\pi l_{13}+i\pi
l_{24}}\right)\left(\frac{1}{2}\z_1\inn\z_4\,\z_2\inn\z_3
-\z_1\inn\z_2\,\z_3\inn\z_4
+\frac{1}{2}\z_1\inn\z_3\,\z_2\inn\z_4\right). \labell{ac01}\eeqa
$A_c^4$ contains terms with four momenta which are non vanishing
even in the $B=0$ limit. They are
 \beqa A_c^4&\!\!=\!\!&
\frac{-icT_p}{6}\left(e^{i\pi l_{12}+i\pi l_{34}} +e^{i\pi
l_{14}-i\pi l_{23}}+e^{i\pi l_{13}-i\pi l_{24}} +e^{i\pi
l_{12}-i\pi l_{34}}+e^{i\pi l_{14}+i\pi l_{23}} +e^{i\pi
l_{13}+i\pi
l_{24}}\right)\nonumber\\
&&\times\left(\frac{}{}\z_1\inn\z_4\,\z_2\inn\z_3\left((k_2 \inn
k_3)(k_1\inn k_4) -(k_1\inn k_2)(k_3\inn k_4)-(k_1\inn
k_3)(k_2\inn k_4)\frac{}{}\right)\right.
\nonumber\\
&&+\z_1\inn\z_2\,\z_3\inn\z_4\left((k_1\inn k_2)(k_3\inn
k_4)-(k_2\inn k_3)(k_1\inn k_4)-(k_1\inn k_3)(k_2\inn
k_4)\frac{}{}\right)
\nonumber\\
&&\left. +\z_1\inn\z_3\,\z_2\inn\z_4\left((k_1\inn k_3)(k_2\inn
k_4) -(k_1\inn k_2)(k_3\inn k_4)-(k_2\inn k_3)(k_1\inn
k_2)\frac{}{}\right)\frac{}{}\right).\labell{ac41}\eeqa And
$A_c^n$ with $n>4$ contains contact terms with more than four
momenta. We refer  readers to \cite{mrg1} for comparing  the
contact terms $A_c^0$ and $A_c^4$ with the noncommutative DBI
action. Here we consider only the simple case that the background
field $B=0$. In this case, $A_c^0=0$, the phase factors in $A_c^4$
reduce to number 6 and $c\rightarrow 1$. Expanding the square
root of determinant, one finds that the tachyonic DBI action has
the following four scalars coupling:\beqa S&=&-T_p\int
d^{p+1}x\,V(T)\sqrt{-\det(\eta_{ab}+\prt_aX^i\prt_bX_i+
2\pi\alpha'\prt_aT\prt_bT)}\,\,,\labell{biact}\\
&=&-T_p\int d^{p+1}x
\left(-\frac{1}{4}(\prt_aX^i\prt_bX_i)(\prt^bX_j\prt^aX^j)+
\frac{1}{8}(\prt_aX^i\prt^aX_i)^2+\cdots\right)\,\,.
\nonumber\eeqa It is easy to verify that couplings above exactly
reproduce the contact terms of $A_c^4$ when $B=0$. Note that in
reaching to this result, one does not need to use the on-shell
conditions for external states. This confirms that the off-shell
amplitude \reef{atoff} is consistent with the DBI action.

\section{Two scalars and two tachyons amplitude} The amplitude
describing scattering of two tachyons to two scalars on the
world-volume of a non-commutative non-BPS D$_p$-brane is
evaluated in \cite{mrg1}. This amplitude has only massless pole in
the $t$-channel , that is
 \beqa A(\z_2,\z_3)\,=\,A_t(\z_2,\z_3)&= &
-\frac{icT_p}{\pi\alpha'}\z_2\inn\z_3
\left(\frac{}{}\frac{\Ga(-2t)\Ga(1/2- 2s)}
 {\Ga(-1/2-2t-2s)}(e^{i\pi
l_{12}+i\pi l_{34}}
+e^{i\pi l_{14}-i\pi l_{23}})\right.\nonumber\\
&&-\frac{\Ga(1/2-2s)\Ga(1/2-2u)}{\Ga(-2s-2u)}(e^{i\pi l_{13}-i\pi
l_{24}}+e^{i\pi
l_{12}-i\pi l_{34}})\nonumber\\
&&\left.+\frac{\Ga(-2t)\Ga(1/2-2u)}{\Ga(-1/2-2t-2u)}(e^{i\pi
l_{14}+i\pi l_{23}}+e^{i\pi l_{13}+i\pi l_{24}}) \frac{}{}
\right)\,\,,\labell{atone}\eeqa where we have normalized the
amplitude here by the factor $icT_p/(\pi\alpha')$. The Mandelstam
variables $s,t,u$ defined in \reef{mandel}, and they satisfy the
on-shell condition $s+t+u=-1/2$.

Using  relation $s+t+u=-1/2$ in \reef{atone}, one can write the
gamma functions in \reef{atone} as those appearing in
\reef{gamma1}, \ie \beqa
\frac{\Ga(-2t)\Ga(1/2-2s)}{\Ga(-1/2-2t-2s)}&=&
\frac{\Ga(-2t)\Ga(1+t+u-s)}{\Ga(u-s-t)}\,\,,\nonumber\\
\frac{\Ga(1/2-2s)\Ga(1/2-2u)}{\Ga(-2s-2u)}&=&
\frac{\Ga(1+t+u-s)\Ga(1+s+t-u)}{\Ga(1+2t)}\,\,,\nonumber\\
\frac{\Ga(-2t)\Ga(1/2-2u)}{\Ga(-1/2-2t-2u)}&=&
\frac{\Ga(-2t)\Ga(1+t+s-u)}{\Ga(s-u-t)}\,\,.\labell{gamma2}\eeqa
Using these form for the gamma  functions in \reef{atone}, one
can see that the on-shell amplitudes \reef{atone} and \reef{aton}
can be written in exactly the same form. Now using the guiding
principle the the off-shell amplitude should also be similar, we
propose the following off-shell extension for the amplitude
\reef{atone}: \beqa A_t^{{\rm off}}
&=&-\frac{icT_p}{\pi\alpha'}\z_2\inn\z_3\left(\left(\frac{}{}
\frac{\Ga(-2t)\Ga(1+t+u-s)}{\Ga(u-s-t)}+F-F^{{\rm on}}
+f'\right)(e^{i\pi l_{12}+i\pi l_{34}}
+e^{i\pi l_{14}-i\pi l_{23}})\right.\nonumber\\
&&-\left(\frac{\Ga(1+t+u-s)\Ga(1+s+t-u)}{\Ga(1+2t)}-F+F^{{\rm on}}
-g'\right) (e^{i\pi l_{13}-i\pi l_{24}}+e^{i\pi
l_{12}-i\pi l_{34}})\nonumber\\
&&\left.+\left(\frac{\Ga(-2t)\Ga(1+t+s-u)}{\Ga(s-u-t)}+F-F^{{\rm
on}} +h'\right) (e^{i\pi l_{14}+i\pi l_{23}}+e^{i\pi l_{13}+i\pi
l_{24}}) \frac{}{} \right)\,\,,\labell{atoff1}\eeqa where $F$ is
given by \reef{F} in which $k_2,k_3$ are momenta of the scalar
states and $k_1,k_4$ are momenta of the tachyons. $F^{{\rm on}}$
means function $F$ in which the momenta are on-shell. It has the
following value: \beqa F^{{\rm
on}}&=&-\frac{\pi^2}{6}\alpha'k_2\inn k_3\,\,.\eeqa  Functions
$f'(k_1,k_2,k_3,k_4),g'(k_1,k_2,k_3,k_4),h'(k_1,k_2,k_3,k_4)$ in
\reef{atoff1} must be zero on-shell. We expect them to be related
to the higher derivative terms in the tachyonic DBI action in
which we are not interested in this paper. So from now on we
ignore them.  The Madelstam variables in the off-shell amplitude
\reef{atoff1} satisfy \reef{mandel1}. Obviously the amplitude
\reef{atoff1} has expected infinite tower of poles in the
$t$-channel, and reduces to \reef{atone} upon imposing the
on-shell conditions.

 Using
the low energy expansion \reef{gamma} for the gamma functions in
\reef{atoff1} one finds massless pole and some contact terms at
each order of $\alpha'$. Now in  field theory,  since the kinetic
term of tachyon in the tachyonic DBI action \reef{biact} is
exactly like the kinetic term of the massless scalar field, with
different normalization \ie $X^i\rightarrow
\sqrt{2\pi\alpha'}T$,  the massless $t$-channel pole for
scattering two scalars to two tachyons is exactly like the
off-shell amplitude \reef{atfield} in which
$\z_1\inn\z_4\rightarrow 2\pi\alpha'$. This pole is exactly the
massless pole of the string theory amplitude \reef{atoff1}.
Subtracting the massless pole of \reef{atoff1}, one finds the
following contact terms:
  \beqa
A_c^{{\rm off}}
(\z_2,\z_3)&=&A_c^0(\z_2,\z_3)+A_c^4(\z_2,\z_3)+\cdots\,\,,
\label{ac}\eeqa where $A_c^0$  includes the contact terms that
have no momentum and is zero when the background field vanishes,
\beqa
A_c^0&=&-\frac{icT_p}{\pi\alpha'}\z_2\inn\z_3\left(\frac{1}{2}\left(e^
{i\pi l_{12}+i\pi l_{34}} +e^{i\pi l_{14}-i\pi
l_{23}}\right)\right.\nonumber\\
&&\left.-\left(e^{i\pi l_{13}-i\pi l_{24}} +e^{i\pi l_{12}-i\pi
l_{34}}\right)+\frac{1}{2}\left(e^{i\pi l_{14}+i\pi l_{23}}
+e^{i\pi l_{13}+i\pi l_{24}}\right)\right)\,\,.\labell{ac02} \eeqa
$A_c^4$ includes contact terms that have at most four momenta and
is non-zero even when $B=0$, \ie
 \beqa
A_c^4
 &\!\!\!=\!\!\!&
\frac{-4i\pi cT_p}{3\alpha'}\left(e^{i\pi l_{12}+i\pi l_{34}}
+e^{i\pi l_{14}-i\pi l_{23}}+e^{i\pi l_{13}-i\pi l_{24}} +e^{i\pi
l_{12}-i\pi l_{34}}+e^{i\pi l_{14}+i\pi l_{23}} +e^{i\pi
l_{13}+i\pi
l_{24}}\right)\nonumber\\
&&\times\z_2\inn\z_3\left(\frac{\alpha'^2}{4}(k_2\inn
k_3)(k_1\inn k_4)-\frac{\alpha'^2}{4}(k_1\inn k_2)(k_3\inn
k_4)-\frac{\alpha'^2}{4}(k_1\inn k_3)(k_2\inn
k_4)+\frac{\alpha'}{8}(k_2\inn k_3) \right)\labell{ac42}\eeqa The
dots in \reef{ac} are the $O(\alpha'^3)$ terms of the gamma
function expansion and the $f',g',h'$ functions in \reef{atoff1}.
Again we refer readers to \cite{mrg1} for comparing  the contact
terms $A_c^0$ and $A_c^4$ above with the noncommutative DBI
action. Here we consider only the simple case that the background
field $B=0$. The tachyonic DBI action \reef{biact} has the
following
 terms: \beqa
{\cal{L}}(X,X,T,T)&=&-T_p \left(-\frac{\pi}{4}T^2
(\prt_aX^i\prt^aX_i)
\right.\labell{lpptt}\\
&&\left.-\pi\alpha'(\prt_aX^i\prt_bX_i)(\prt^bT\prt^aT)+
\frac{\pi\alpha'}{2}(\prt_aX^i\prt_aX_i)(\prt_bT\prt^bT)\right)\,\,,
\nonumber\eeqa It is easy to see  that the coupling in the first
line produce the term in \reef{ac42} which has two momenta, and
the couplings in the second line above reproduce the other terms
in \reef{ac42}. This confirms that the off-shell amplitude
\reef{atoff1} is consistent with the tachyonic DBI action, and the
expectation that the function $f',g',h'$ are related to the
higher derivative terms in this action, \ie their $\alpha'$
expansion does not have constant and term proportional to
$\alpha'$.

\section{Four tachyons amplitude} The amplitude describing
scattering of two tachyons to two other tachyons on the
world-volume of a non-commutative non-BPS D$_p$-brane is
evaluated in \cite{mrg1}. This amplitude has massless pole in all
$s$-, $t$- and $u$-channels, that is, \beqa A&=&\,A_s+\,A_u+
\,A_t\,\,.\nonumber\eeqa However, in this case the amplitudes in
all channels are identical \ie $A_s=A_t=A_u$, and
 \beqa
A_t&=&-2icT_p\left(\frac{}{}
\frac{\Ga(-2t)\Ga(-2s)}{\Ga(-1-2s-2t)}(e^{i\pi l_{12}+i\pi l_{34}}
+e^{i\pi l_{14}-i\pi l_{23}})\right.\nonumber\\
&&-\frac{\Ga(-2s)\Ga(-2u)}{\Ga(-1-2s-2u)}(e^{i\pi l_{13}-i\pi
l_{24}}+e^{i\pi
l_{12}-i\pi l_{34}})\nonumber\\
&&\left.+\frac{\Ga(-2t)\Ga(-2u)}{\Ga(-1-2t-2u)}(e^{i\pi
l_{14}+i\pi l_{23}}+e^{i\pi l_{13}+i\pi l_{24}}) \frac{}{}
\right)\,\,,\labell{atttt}\eeqa where we have normalized the
amplitude by the factor $2icT_p$. The variables
 $s,t,u$ satisfy the on-shell condition $ s+t+u=-1$.

Again using  relation $s+t+u=-1$ in \reef{atttt}, one can write
the gamma functions in it as those appearing in \reef{gamma1}, \ie
\beqa \frac{\Ga(-2t)\Ga(-2s)}{\Ga(-1-2t-2s)}&=&
\frac{\Ga(-2t)\Ga(1+t+u-s)}{\Ga(u-s-t)}\,\,,\nonumber\\
\frac{\Ga(-2s)\Ga(-2u)}{\Ga(-1-2s-2u)}&=&
\frac{\Ga(1+t+u-s)\Ga(1+s+t-u)}{\Ga(1+2t)}\,\,,\nonumber\\
\frac{\Ga(-2t)\Ga(-2u)}{\Ga(-1-2t-2u)}&=&
\frac{\Ga(-2t)\Ga(1+t+s-u)}{\Ga(s-u-t)}\,\,.\nonumber\eeqa Using
these form for the gamma  functions in \reef{atttt},
\reef{gamma2} for the gamma functions in \reef{atone} and
\reef{gamma1} for the gamma functions in \reef{aton}, one can
easily see that the on-shell amplitudes \reef{atttt},
\reef{atone} and \reef{aton} can be written in exactly the same
form. Now using the guiding principle that the off-shell
amplitudes should also be similar, we propose the following
off-shell extension for the amplitude \reef{atttt}: \beqa
A_t^{{\rm off}}&=&-2icT_p\left(\left(\frac{}{}
\frac{\Ga(-2t)\Ga(1+t+u-s)}{\Ga(u-s-t)}+F-F^{{\rm on}}
+f''\right)(e^{i\pi l_{12}+i\pi l_{34}}
+e^{i\pi l_{14}-i\pi l_{23}})\right.\nonumber\\
&&-\left(\frac{\Ga(1+t+u-s)\Ga(1+s+t-u)}{\Ga(1+2t)}-F+F^{{\rm on}}
-g''\right) (e^{i\pi l_{13}-i\pi l_{24}}+e^{i\pi
l_{12}-i\pi l_{34}})\nonumber\\
&&\left.+\left(\frac{\Ga(-2t)\Ga(1+t+s-u)}{\Ga(s-u-t)}+F-F^{{\rm
on}} +h''\right) (e^{i\pi l_{14}+i\pi l_{23}}+e^{i\pi l_{13}+i\pi
l_{24}}) \frac{}{} \right)\,\,,\labell{atoff2}\eeqa  where $F$ is
given by \reef{F} in which all the momenta are the tachyon
momenta. $F^{{\rm on}}$ in this case is\beqa F^{{\rm
on}}&=&-\frac{\pi^2}{6}\left(\alpha' k_2\inn k_3+\alpha' k_1\inn
k_4+\frac{1}{2}\right)\,\,.\eeqa Functions
$f''(k_1,k_2,k_3,k_4),g''(k_1,k_2,k_3,k_4),h''(k_1,k_2,k_3,k_4)$
must be zero on-shell. We expect them to be related to the higher
derivative terms in the tachyonic DBI action in which we are not
interested in the present paper. So we ignore them from now on. It
is obvious that the off-shell amplitude \reef{atoff2} has the
expected infinite tower of poles in the $t$-channel, and reduces
to \reef{atttt} upon imposing on-shell conditions.

Using the low energy expansion \reef{gamma} for the gamma
functions in \reef{atoff2}, one finds massless pole and some
contact terms at each order of $\alpha'$. In field theory, the
massless pole is again exactly like \reef{atfield} in which
$\z_2\inn\z_3\z_1\inn\z_4\rightarrow (2\pi\alpha')^2$.
Subtracting it from the off-shell amplitude \reef{atoff2} and
repeating the same analysis for $A_s$ and $A_u$, one finds the
following total contact terms: \beqa A_c^{{\rm off}}&=&
-\frac{8\pi^2icT_p}{3}\nonumber\\
&&\times\left(e^{i\pi l_{12}+i\pi l_{34}} +e^{i\pi l_{14}-i\pi
l_{23}}+e^{i\pi l_{13}-i\pi l_{24}} +e^{i\pi l_{12}-i\pi
l_{34}}+e^{i\pi l_{14}+i\pi l_{23}} +e^{i\pi l_{13}+i\pi
l_{24}}\right)\nonumber\\
&&\times\left(-\frac{\alpha'^2}{4}(k_2\inn k_3)(k_1\inn k_4)
-\frac{\alpha'^2}{4}(k_1\inn k_2)(k_3\inn
k_4)-\frac{\alpha'^2}{4}(k_1\inn k_3)(k_2\inn k_4)\right.\labell{ac43}\\
&&\left.+ \frac{\alpha'}{8}(k_2\inn k_3+k_1\inn k_4+k_1\inn
k_3+k_2\inn k_4+k_1\inn k_2+k_3\inn k_4)+
\frac{3}{16}\right)+\cdots\,,\nonumber\eeqa where dots represent
the $O(\alpha'^3)$ terms of the gamma function  and functions
$f'',g'',h''$.  Here again to compare it with the tachyonic DBI
action for the case that $B=0$, one should replace the phase
factor above by number 6 and $c\rightarrow 1$. The action
\reef{biact} has the following terms: \beqa
{\cal{L}}(T,T,T,T)&=&-T_p\left( \beta
T^4-\frac{\pi^2\alpha'}{2}T^2(\prt_aT\prt^aT)-
\frac{\pi^2\alpha'^2}{2}(\prt_aT\prt^aT)^2\right)\,\,,
\labell{ltttt}\eeqa where  the constant $\beta$ is the
coefficient of $T^4$ in the tachyon potential. One can easily
observe that the term with four derivatives reproduces exactly the
four momentum contact terms in \reef{ac43}, term with two
derivatives reproduces the two momentum contact terms, and the
$T^4$ term produces the constant term in \reef{ac43} provided that
$\beta=\pi^2/8$.

 The tachyon potential expanded around its
maximum, \ie around $T_{max}=0$, has then the following expansion:
\beqa
V(T)=1-\frac{\pi}{2}T^2+\frac{\pi^2}{8}T^4+O(T^6)\,\,.\labell{finalv}
\eeqa On the other hand A. Sen has shown that  the  tachyon
potential in the tachyonic DBI action  has minimum at
$T_{min}\rightarrow \infty$, and the behavior of the potential
around the minimum should be like $e^{-\sqrt{\pi}T}$
 \cite{sen3}\footnote{Note that the convention
in \cite{sen3} sets $\alpha'=1$.}. A  tachyon potential with this
 behavior at  $T_{min}\rightarrow \infty$,
and expansion \reef{finalv} at the maximum of potential has been
suggested in \cite{mrg1}. Our calculation however does not rule
out the possibility of other expansion  for the potential. This
stems from the fact that functions $f'',g'',h''$ in \reef{atoff2}
which must be zero on-shell, might have a constant term in their
$\alpha'$ expansion.  Another potential which has the above
expansion around maximum is the following: \beqa V(T)&=&e^{-\pi
T^2/2}\,\,.\nonumber\eeqa This potential  is consistent with the
sigma model effective action\cite{aat2}.

Extending the similarity between the tachyon and scalar to the
higher derivative terms as well, one might expect that the
functions $f',g',h'$ in \reef{atoff1} to be related to the
functions $f,g,h$ in \reef{atoff} as $f'=f-f^{{\rm on}},
g'=g-g^{{\rm on}}, h'=h-h^{{\rm on}}$, and similarly the
functions $f'',g'',h''$ in \reef{atoff2} as $f''=f-f^{{\rm on}},
g''=g^{{\rm on}},h''=h^{{\rm on}}$. This means that the higher
derivative corrections to the tachyonic DBI action  and  the
higher derivative terms of DBI action have the same structure.
Similar speculation, in other context, has  been also made in
\cite{sendbi}

Finally to compare the low energy expansion of the off-shell
amplitudes considered in this paper with  the on-shell amplitudes
considered in \cite{mrg1}, one should not that in the off-shell
case the Mandelstam variables are all independent. Hence at low
energy they all go to zero, as we have considered in this paper.
Whereas the on-shell relation like \reef{mandel1} constrains  the
variables $s,t,u$ not to be  independent anymore in the on-shell
amplitudes. In this case one should write the amplitudes in terms
of only independent variables then expands  them at the low
energy. For instance, as it has been done in \cite{mrg1}, for
studying $A_t$ at low energy one writes $t$ in terms of $s$ and
$u$. Then expands the amplitude at $\alpha'k_i\inn k_j\rightarrow
0$. In this regards, both the off-shell amplitudes proposed in
this paper  and the on-shell amplitudes gives the same low energy
expansion.

{\bf Acknowledgement}: I would like to thank M. Alishahiha, A. Sen
and A.A. Tseytlin for comments.

%\newpage

\end{document}